\documentclass{IOS-Book-Article}     
\usepackage{amsmath}
\usepackage{amssymb}
\usepackage{amsfonts}
\usepackage{amsthm}
\usepackage{graphicx}
\usepackage{color}
\usepackage{url}
\usepackage{listings}
\usepackage{setspace}
\usepackage{subfigure}
\usepackage{wrapfig}
\usepackage{rotating}
\usepackage{algorithm}
\usepackage{subfigure}
\usepackage{url}
\usepackage{pgf}
\usepackage{tikz}
\usepackage{listings}

\theoremstyle{plain}            

\newtheorem{theorem}{Theorem}[section]

\newtheorem{proposition}[theorem]{Proposition}

\theoremstyle{definition}       

\newtheorem{definition}[theorem]{Definition}

\lstdefinelanguage  {algorithm}
{keywords=[18]{begin,end,if,else,elsif,then,while,for,repeat,do,input,output,proc,return,let,with,where,set},
morecomment=[s]{\{}{\}} }
\lstnewenvironment {pseudocode}{\lstset{tabsize=4, language=algorithm,
texcl=true, mathescape=true, keywordstyle=\color{black}\bf,
basicstyle=\footnotesize\color{darkgray}, commentstyle=\itshape,columns=flexible }
\lstset{moredelim=[is][\underline]{_}{_}}
}{}

\begin{document}
\begin{frontmatter}          
\title{Computing alignment plots efficiently}
\runningtitle{Computing alignment plots efficiently}

 \author{Peter Krusche and Alexander Tiskin*},
\runningauthor{P. Krusche and A. Tiskin}
\address{Dept. of Computer Science, University of Warwick, Coventry,
CV4~7AL, UK}
\begin{abstract}
Dot plots are a standard method for local comparison of biological
sequences. In a dot plot, a substring to substring distance is computed for all
pairs of fixed-size windows in the input strings. Commonly, the Hamming distance
is used since it can be computed in linear time. However, the Hamming distance
is a rather crude measure of string similarity, and using an alignment-based edit
distance can greatly improve the sensitivity of the dot plot method.
In this paper, we show how to compute alignment plots of the latter type
efficiently. Given two strings of length $m$ and $n$ and a window size $w$, this
problem consists in computing the edit distance between all pairs of
substrings of length $w$, one from each input string. The problem can be
solved by repeated application of the standard dynamic programming
algorithm in time $O(mnw^2)$. This paper gives an improved
data-parallel algorithm, running in time $O(mnw/\gamma/p)$ using vector
operations that work on $\gamma$ values in parallel and $p$ processors.
\end{abstract}


\end{frontmatter}


\section{Introduction}

\let\thefootnote\relax\footnotetext{* Research supported by the
 Centre for Discrete Mathematics and Its Applications
 (DIMAP), University of Warwick, EPSRC award EP/D063191/1.
 Computational resources were provided by the Centre for Scientific
 Computing at the University of Warwick.
 }
Dot plots are a standard method for local comparison of two biological
sequences introduced by Gibbs/McIntyre~\cite{Gibbs:70} and
Maizel/Lenk~\cite{Maizel_Lenk:81}.
When creating a dot plot, a substring to substring distance is computed
for all pairs of fixed-size windows in the input strings. The result
can be visualized by a plot showing a dot for each pair of windows that
achieves a distance below a fixed threshold.
Commonly, the Hamming distance 
is used~\cite{Maizel_Lenk:81,Krumsiek:07}, since it can be
computed very efficiently. 
However, the Hamming distance is a rather crude measure of string similarity.
Using a string edit distance or alignment score (see e.g.\ \cite{Gusfield:97})
for dot plot filtering can greatly improve the sensitivity of the method.
In the context of biological sequence comparison, this idea has been implemented
by Ott et al.~\cite{Ott:09}, where a sequential algorithm is given which
creates an alignment plot for two strings of lengths $m$ and $n$ using a 
fixed window length~$w$ 
in time $O(m n w^2)$. 

In this paper, we give an improved data-parallel algorithm, running in time
$O(mnw/\gamma)$ using vector operations that work on $\gamma$ values in
parallel, and show experimental speedups from an implementation using
MMX~\cite{Intel:09}. Furthermore, we demonstrate that the algorithm can be
parallelized to run on multiple processors using MPI~\cite{Snir+:95}.

\section{Computing longest common subsequences and string
alignments}\label{sec:lcs} 
Let $x = x_1 x_2 \ldots x_m$ and $y = y_1 y_2 \ldots
y_n$ be two strings over an alphabet $\Sigma$ of size $\sigma$. We distinguish between contiguous
\textit{substrings} of a string $x$, which can be obtained by removing zero or
more characters from the beginning and/or the end of $x$, and \textit{subsequences},
which can be obtained by deleting zero or more characters in any position.
The \textit{longest common subsequence} (LCS) of two strings is the longest
string that is a subsequence of both input strings; its length (the LLCS) is a
measure for the similarity of the two strings.
Substrings of length $w$ are called \textit{$w$-windows}.
For a given $w$, the length of the LCS of two $w$-windows $x_i \ldots x_{i+w-1}$
and $y_j \ldots y_{j+w-1}$ will be denoted as $\mathit{WLCS}(i,j)$. 
An \textit{alignment plot} for $x$ and $y$ consists of all values
$\mathit{WLCS}(i,j)$ with $i \in \{1, 2, \ldots, m-w\}$, $j \in \{1, 2, \ldots, n-w \}$.

Although the LCS is more accurate than the Hamming score, 
more general similarity measures are of interest in practice. A standard
interpretation of LCS is \textit{string alignment}~\cite[p. 209 ff.]{Gusfield:97}. 
An alignment of strings $x$ and $y$ is obtained by putting a subsequence of $x$
into one-to-one correspondence with a (not necessarily identical) subsequence
of $y$, character by character and respecting the index order. The
corresponding pairs of characters, one from $x$ and the other from $y$, are
said to be \textit{aligned}. A character not aligned with a character of
another string is said to be aligned with a \textit{gap} in that string.
Finding the LCS corresponds to computing a maximum alignment when assigning the scores
 $w_{=} = 1$ to aligning a matching pair of characters, $w_{\text{\textvisiblespace}} = 0$ to
inserting a gap, and $w_{\neq} = 0$ to aligning two mismatching characters.
More general alignments than the LCS can be obtained using the standard dynamic
programming algorithm~\cite{Wagner+:74,Needleman1}, which allows for gap
penalties as well as different scores for each individual pair of matching/mismatching characters,
forming a \textit{pairwise score matrix}. Any algorithm for LCS computation can
be generalized to pairwise score matrices with small rational scores at the
price of a constant factor blow-up of the input
strings~\cite{TiskinL:07:Applications}. In this paper, we will consider
alignments with match score $w_{=} = 1$, mismatch score $w_{\neq} = 0$ and gap penalty
$w_{\text{\textvisiblespace}} = -0.5$.
To compute these alignments, we modify the input strings by adding a new character
\texttt{\$} to the alphabet, which we insert before every character in both
input strings such that e.g.\ \texttt{abab} transforms into \texttt{\$a\$b\$a\$b}. 
 For input strings $x$ and $y$ of length $m$ and $n$, the alignment score
 $S(x,y)$ can be retrieved from LLCS of the modified strings $x'$ and $y'$ as
 $S(x,y) = LLCS(x',y') - 0.5 \cdot (m + n)$.
We expect the running time of the seaweed
algorithm to increase by a factor of four by this reduction, as both input
strings double in size.

Our new algorithms are based on semi-local sequence alignment~\cite{Tiskin:05},
for which we now give the necessary definitions. Throughout this paper, we will
denote the set of integers $\{i, i+1, \ldots, j\}$ by $[i : j]$, and the set of odd half-integers
$\{i+\frac{1}{2}, i+\frac{3}{2}, \ldots, j - \frac{1}{2}\}$ by $\langle i : j
\rangle$. We will further mark odd half-integer variables by a ``$\hat{\ }$''
symbol. When indexing a matrix $M$ by odd-half integer values $\hat{\imath}$
and $\hat{\jmath}$, we define that $M(\hat{\imath}, \hat{\jmath}) = M(i, j)$
with $i = \hat{\imath}-1/2$ and $j = \hat{\jmath} - 1/2$. Therefore, if a
matrix has integer indices $[1:m]\times [1:n]$, it has odd-half integer
indices $\langle 1:m+1\rangle\times \langle 1:n+1\rangle$. 
We also define the \textit{distribution matrix} $D^\Sigma$ of an $m\times n$
matrix $D$ as $ D^\Sigma(i,j) = \sum
D(\hat{\imath},\hat{\jmath})$ with $(\hat{\imath},\hat{\jmath})\in
\langle i:m+1 \rangle \times \langle 1:j \rangle
\text{.}$

Let the \textit{alignment dag (directed acyclic graph)} $G_{x,y}$ for two strings $x$
and $y$ be defined by a set of vertices $v_{i,j}$ with
$i \in [0:m]$ and $j \in [0:n]$ and edges as follows.
We have horizontal and vertical edges $v_{i, j-1} \rightarrow v_{i, j}$ and
$v_{i-1, j} \rightarrow v_{i, j}$ of score~0. Further, we introduce diagonal edges
$v_{i-1, j-1} \rightarrow v_{i, j}$ of score~1, which are present only if
$x_i = y_j$.
Longest common subsequences of a substring $x_i x_{i+1} \ldots x_{j}$ and $y$
correspond to highest-scoring paths in this graph from $v_{i-1,0}$ to $v_{j,m}$.
When drawing the 
alignment dag in the plane, its horizontal and vertical
edges partition the plane into rectangular \textit{cells} each of which,
depending on the input strings, may contain a diagonal edge or not. For every pair of
characters $x_i$ and $y_j$, we define a corresponding cell
$(i-\frac{1}{2}, j-\frac{1}{2})$. Cells corresponding to a matching pair of
characters are called \textit{match cells}, and cells corresponding to
mismatching characters are called \textit{mismatch cells}.

Solutions to the \textit{semi-local LCS problem} are given by a
\textit{highest-score matrix} which we define as follows.
%
In a highest-score matrix $A_{x,y}$, each entry $A_{x,y}(i,j)$ is defined as
the length of the highest-scoring path in $G_{x,y}$ from $v_{i-1,0}$ to $v_{j,m}$.
Each entry $A_{x,y}(i,j)$ with $0 < i < j < n$ gives the LLCS
of $x$ and substring $y_i \ldots y_j$. 

Since the values of
$A_{x,y}(i,j)$ for different $i$ and $j$ are strongly correlated, it is possible to derive an implicit, space-efficient representation
of matrix $A_{x,y}(i,j)$. This implicit representation of a semi-local highest-score
matrix consists of a set of \textit{critical points}.
The critical points of a highest-score matrix $A$ are defined as the set of odd
half-integer pairs $(\hat{\imath}, \hat{\jmath})$ such that
$A(\hat{\imath}+\frac{1}{2},\hat{\jmath}-\frac{1}{2}) + 1 =
 A(\hat{\imath}-\frac{1}{2},\hat{\jmath}-\frac{1}{2}) =
 A(\hat{\imath}+\frac{1}{2},\hat{\jmath}+\frac{1}{2}) =
 A(\hat{\imath}-\frac{1}{2},\hat{\jmath}+\frac{1}{2})
 $.
Consider a highest-score matrix $A$. The matrix $D_A$ with $D_A(\hat{\imath},
\hat{\jmath}) = 1$ if $(\hat{\imath},\hat{\jmath})$ is a critical point in $A$, 
and $D_A(\hat{\imath},\hat{\jmath}) = 0$ otherwise, is called the 
\textit{implicit highest-score matrix}.

Tiskin~\cite{Tiskin:05} showed that in order to represent a highest-score matrix
for two strings of lengths $m$ and $n$, exactly $m+n$ such critical points are sufficient.
\begin{theorem}[see~\cite{Tiskin:05} for proof]\label{thm:implicitrep}
A highest-score matrix $A$ can be represented implicitly using only 
$O(m+n)$ space by its implicit highest-score matrix $D_A$, which 
is a permutation matrix. We have: $A(i,j) = j - i -
D^\Sigma_A(i,j)$, where $D_A^\Sigma$ is the distribution matrix of the
implicit highest-score matrix $D_A$. 
\end{theorem}
The set of critical points can be obtained using the \textit{seaweed algorithm}
(by Alves et al.~\cite{Alves+:08}, based on Schmidt~\cite{Schmidt:98}, adapted by
Tiskin~\cite{Tiskin:05}) which computes critical points by dynamic programming
on all prefixes of the input strings.
This method is graphically illustrated by tracing \emph{seaweeds} that
start at odd half-integer positions between two adjacent vertices $v_{0, \hat{\imath} - \frac{1}{2}}$ and
$v_{0,\hat{\imath} + \frac{1}{2}}$ in the top row of the alignment dag,
and end between two adjacent vertices $v_{m, \hat{\jmath} - \frac{1}{2}}$ and
$v_{m,\hat{\jmath} + \frac{1}{2}}$ in the bottom row. Each critical point is
computed as the pair of horizontal start and end coordinates of such a seaweed
(see Algorithm~\ref{alg:seaweed_algorithm}). Two seaweeds enter every cell in
the alignment dag, one at  the left and one at the top. The seaweeds
proceed through the cell  either downwards or rightwards. In the cell, the
directions of these seaweeds are interchanged either if there is a match
$x_{k} = y_{l}$, or if the same  pair of seaweeds have already crossed.
Otherwise, their directions remain unchanged and the seaweeds cross. 
%
By Theorem~\ref{thm:implicitrep}, the length of the 
highest-scoring path in $G_{x,y}$ from $v_{i-1,0}$ to $v_{j,m}$
can be computed by counting the number of seaweeds which both start and end
within $\langle i:j\rangle$. 

\begin{algorithm}[b]
\caption{\label{alg:seaweed_algorithm}The Seaweed Algorithm}
\begin{minipage}[c]{0.5\textwidth}

\vspace{.1cm}
\includegraphics[height=4cm,clip]{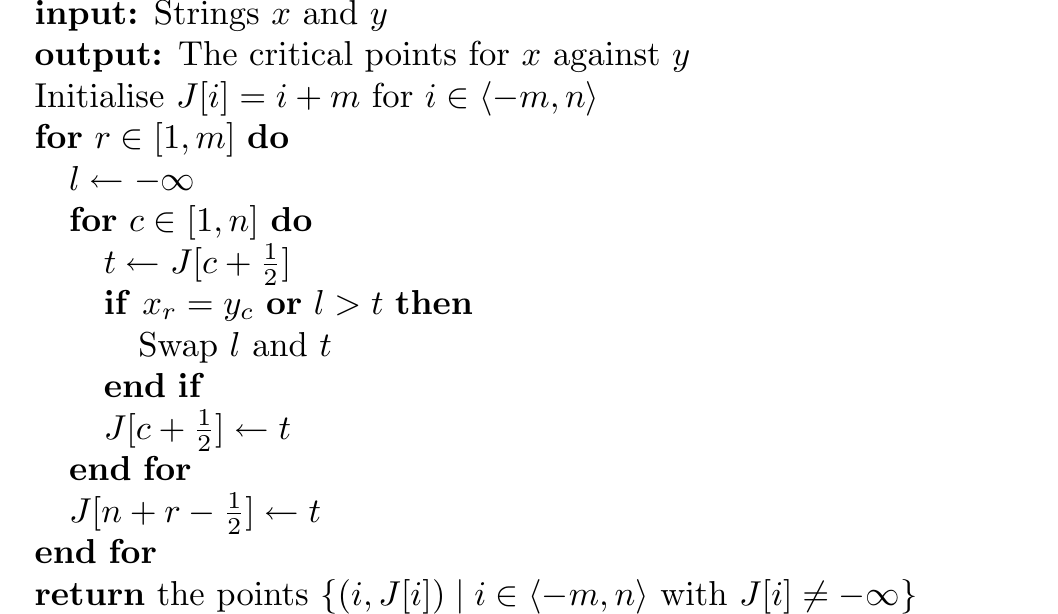}

\vspace{.1cm}
\end{minipage}
\begin{minipage}[c]{0.49\textwidth}

\vspace{.1cm}
\begin{center}
\includegraphics[height=4cm,clip]{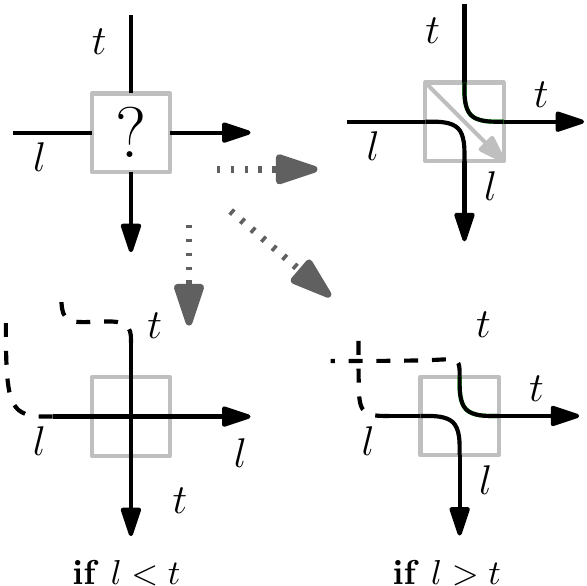}
\end{center}

\vspace{.1cm}
\end{minipage}
\end{algorithm}



\section{Data-parallel window-window comparison using
seaweeds}\label{sec:seawinwin} 
The seaweed algorithm can be used to compute the LCS
of all pairs of $w$-windows simultaneously in time $O(wn)$ for two strings of
respective lengths $w$ and $n$ (i.e.\ one of the strings consists of only one
$w$-window). By Theorem~\ref{thm:implicitrep}, the LCS of $x$ and any
$w$-window $y_i \ldots y_{i+w-1}$ is computed as the number of seaweeds starting
and ending within the odd half-integer range $\langle i:i+w \rangle$. 
By keeping track of only these seaweeds in a sliding window,
our algorithm can compute the LLCS for all $w$-windows in a single pass over all columns of cells in the
alignment dag.
We obtain an improved algorithm for comparing all pairs of
$w$-windows.
\begin{theorem} \label{thm:basic_wlcs}
Given two strings $x$ and $y$ of lengths $m$ and $n$, the LLCS
for all pairs of $w$-windows between $x$ and $y$ can be computed in time
$O(mnw)$.
\end{theorem}
\begin{proof}
We apply the seaweed algorithm for computing the implicit $(w,1)$-restricted
highest-score matrices for $y$ and all substrings of $x$ that have length $w$.
Each application of the seaweed algorithm therefore runs on a
strip of height $w$ and width $n$ of the alignment dag corresponding to $x_i \ldots x_{i+w-1}$
and  $y$. A column of cells in this strip can be processed in time $O(w)$. 
In each column $j$, exactly one new seaweed starts at the top
of the alignment dag, and exactly one seaweed ends at the bottom.
We track seaweeds ending within $\langle j-w:j\rangle$. 
To count the seaweeds that have reached
the bottom of the alignment dag, we maintain a priority queue ${B}$. In each
step, one seaweed reaches the bottom of the alignment dag. 
Furthermore, in each step, we have to delete at most one seaweed from $B$. We
use a priority queue of $\lceil\log n\rceil$-bit integers to represent $B$.
For each seaweed that reaches the bottom, we compute its
starting point and add it to the queue. We delete the minimum
value from the queue if it is smaller than the starting position of the
current $w$-window in string $y$. By using a min-heap~\cite{Cormen+01}, both
operations can be implemented in $O(\log w)$. The number $d$ of
seaweeds which start within $\langle j-w:j\rangle$ is then given by the size of
queue~$B$. The LLCS of $y_{j-w+1}\ldots y_j$ and
$x_i\ldots x_{i+w-1}$ can then be calculated as $w-d$. In total, we have to
process $n$ columns using time $O(w)$ in every strip. Overall, $m-w$ strips
exist, therefore we obtain running time $O(m n w)$.
\end{proof}
While this direct application of the seaweed method gives an asymptotic improvement
on the method of computing the LCS independently for every pair of windows by dynamic
programming, it is not necessarily more practical. The dynamic programming method
can exploit the fact that we are only interested in windows with an alignment score
above a given threshold. More importantly, the dynamic programming method allows
one to improve performance by introducing a step size $h$, and only comparing
$w$-windows starting at positions that are multiples of $h$. We will now show
how to improve the practical performance of the algorithm.

Algorithm~\ref{alg:seaweed_algorithm} requires $O(\log (m+n))$ bits to
represent the start and endpoints of a single seaweed. We first show 
that for computing alignment plots with a fixed window length $w$, $O(\log w)$ bits are
sufficient for tracing seaweeds, independently of the size of the input strings.
To show this, we define the \textit{span} of a seaweed as the horizontal
distance it covers in the alignment dag. A seaweed corresponding to a critical point
$(\hat{\imath}, \hat{\jmath})$ has span $\hat{\jmath} - \hat{\imath}$. Seaweeds
that have a span greater than the window length $w$ are not relevant for
computing the alignment plot, since they will not start and end within a single
window. Furthermore, we are only interested in values with index pairs $(i,j)$
having $i\ \mathrm{mod}\ r = j\ \mathrm{mod}\ r = 0$, where $r$ is the
constant blowup induced by the alignment score (for the scoring
scheme described in the previous section, we have $r=2$). This is equivalent to
computing the semi-local LCS for substrings restricted to length~$w$, starting and ending 
at positions $0\ \mathrm{mod}\ r$.
\begin{definition}
\label{def:restricted_hsm}
Let $A$ be a highest-score matrix. 
The \textit{$(w,r)$-restricted highest-score matrix} $A^{w,r}$ is defined as
$A^{w,r}(i,j)= 
            	A(i,j) 
            	\ \mathrm{if}\ j - i \leq w$ and $i\ \mathrm{mod}\ r = j\
            	\mathrm{mod}\ r = 0$, and  
            	$A^{w,r}(i,j) = \textit{undefined}$ otherwise.
\end{definition}
This restriction on the highest-score matrices allows us to reduce the number of
critical points we need to store, and also to reduce the number of bits required
to represent seaweeds in our computation. 

\begin{proposition}
 To represent a $(w,r)$-restricted highest-score matrix implicitly, we 
 only need to store the
 critical points $(\hat{\imath}, \hat{\jmath})$ of the corresponding unrestricted highest
 score matrix for which $\hat{\jmath}-\hat{\imath} < w$.
\end{proposition}
\begin{proof}
Straightforward from Theorem~\ref{thm:implicitrep} and
Definition~\ref{def:restricted_hsm}.
\end{proof}

\begin{proposition}\label{prop:w_r_bits}
 We can represent a single critical point in a $(w,r)$-restricted
 highest-score matrix for comparing strings $x$ and $y$ using $O(\log(w/r))$ bits.
\end{proposition}
\begin{proof}
 We store the seaweeds in a vector $S$ of size $m+n$, where each vector element
 stores $\lceil \log_2(w/r + 1) \rceil$ bits. For each critical point
 $(\hat{\imath}, \hat{\jmath})$, we have one vector element
 $S(\hat{\imath}+1/2) = \min(2^{w/r+1}-1,(\hat{\jmath}-\hat{\imath})/r)$. Each
 vector element stores the span of the seaweed starting at $\hat{\imath}$.

 It is straightforward to see that we only need $O(\log w)$ bits for a vector
 element: seaweeds in a $(w,r)$-restricted highest-score matrix become irrelevant once
 their span is larger than $w$, since these critical points will not affect any
 LCS for a substring of length $w$ (see Theorem~\ref{thm:implicitrep}).
 In order to reduce the number of bits  to $O(\log w/r)$, we use the fact that we
 only need to answer LCS queries correctly if $i\ \mathrm{mod}\ r = j\
 \mathrm{mod}\ r = 0$. We therefore do not need to distinguish the individual
 $r$ seaweeds starting within $[k:k+r-1]$ with $k \bmod r = 0$: once
 they reach the bottom, we only need to know their starting position within a window of
 size $r$. We can therefore divide the distance values by $r$, which gives
 the claimed number of required bits.
\end{proof}

We now show how to use vector instructions for improving
Algorithm~\ref{alg:seaweed_algorithm} to trace multiple seaweeds in
parallel. 
A practical example for vector parallelism are Intel's MMX
instructions~\cite{Intel:09} for integer vector
arithmetic and comparison (it
would also be possible to implement our algorithm using floating point vector
processing, e.g.\ using SSE~\cite{Intel:09}).
In our algorithm, we assume that all elements $V(j)$ in a vector $V$ are $v$-bit
values. If an element of a vector has all bits set, then this represents the
value of $+\infty$, having $\mathit{INF} \equiv 2^v - 1$.
When carrying out the seaweed algorithm on columns of the alignment dag, 
the result of every comparison in a cell of the column depends on the
comparison result from the cell above it. To be able to process multiple cells
in parallel, we need to process cells by antidiagonals. 
We can then use vector operations to implement each step in the the seaweed
algorithm, as each cell can be processed only using data computed in the previous step.
We need to track seaweeds only if they are within the window of interest. In
order to keep the required value of bits per seaweed as small as possible (and
hence allow a high degree of vector parallelism), we identify seaweeds by the distance of their
starting points to the current column. This distance can be represented using
$v = \lceil \log_2 2w+1 \rceil$ bits (see Proposition~\ref{prop:w_r_bits}).
When advancing to the next column, we use saturated addition to
increment all distances in parallel
(i.e.\ saturated addition of one to vector element ${V}(k)$ gives ${V}(k) + 1$
if $V(k) < 2w$, and $\mathit{INF}$ otherwise).
In each step, we compare the characters corresponding to the cells in the
current antidiagonal using vectorised mask generation. Given two vectors ${V}$
and ${W}$, we generate a mask vector which contains the value $\mathit{INF}$ at
all positions $k$, where ${V}(k) = {W}(k)$ and zero otherwise. 
The seaweed behaviour in mismatch cells is implemented by a compare/exchange
operation which, given two vectors ${V}$ and ${W}$, exchanges ${V}(k)$ and
${W}(k)$ only if ${V}(k) > {W}(k)$. 
To combine the results from the mismatch cells and the match cells, we
require an operation to exchange vector elements conditionally using the mask
vector $M$ generated earlier. Given two vectors ${V}$
and ${W})$, this operation returns vector elements ${V}(k)$ if ${M}(k)
= \mathit{INF}$, and ${W}(k)$ otherwise. 
All these operations can be vectorized efficiently using MMX. Using these
operations, we can implement the seaweed operations from
Algorithm~\ref{alg:seaweed_algorithm} by storing all seaweeds on the current
antidiagonal in a vector $V$ if they enter the respective cell from the left,
and a vector $W$ if they enter the respective cell from the top. We use a
$v$-bit shift operation on vector $W$ in each step to advance the seaweeds
leaving cells at the bottom downwards.
 
\section{Experimental Results}\label{sec:expres}
\begin{table}[b]
\caption{Execution times in seconds and speedups}
\label{tbl:compresults}
\footnotesize
\begin{tabular*}{\textwidth}{l@{}rrrr}
\textbf{Data Set} & \multicolumn{1}{c}{\textbf{Mikey} (2712$\times$628)} &
\multicolumn{1}{c}{\textbf{Berti} (2712$\times$2305)} &
\multicolumn{1}{c}{\textbf{Jimmy} (15k$\times$97k)} &
\multicolumn{1}{c}{\textbf{Henry} (80k$\times$80k)}\\ 
\hline
\hline
\multicolumn{5}{l}{\textit{MMX vector speedup on Linux/x86\_64/1.83GHz
Core2-duo (non-MPI), gcc 4.3.1}}\\
\hline
{Heur} &5.1 ($\div$ 1.0)  &41.1 ($\div$ 1.0) &2677 ($\div$ 1.0)
&11708 ($\div$ 1.0) \\
{BLCS} &3.6 ($\div$ 1.4)  &37.3 ($\div$ 1.1) &3680 ($\div$ 0.7)
&16191 ($\div$ 0.7) \\
{Sea-16} &1.4 ($\div$ 3.6)  &10.8 ($\div$ 3.8) &1026 ($\div$ 2.6)
&4514 ($\div$ 2.6) \\
{Sea-8}  &0.5 ($\div$ 10.2) 	&3.8 ($\div$ 10.8) & 368 ($\div$ 7.3)
&1614 ($\div$ 7.3) \\
\hline
\multicolumn{5}{l}{\textit{Linux desktop system, Core2-quad 2.66GHz, 64-bit,
MPI, gcc 4.3.1}}\\
\hline
1 core  & 0.4 ($\div$ 1.0) & 2.9 ($\div$ 1.0)  & 271 ($\div$ 1.0) & 1199 ($\div$ 1.0) \\ 
4 cores & 0.7 ($\div$ 0.6) & 1.3 ($\div$ 2.2) & 70 ($\div$ 3.9) & 307 ($\div$ 3.9)  \\
\hline
\multicolumn{5}{l}{\textit{IBM HPC Cluster~\cite{website_csc_facilities},
2$\times$dual-core Xeon 3GHz per node, 64-bit, MPI, gcc 4.1.2}}\\
\hline
1 core  & 0.67 ($\div$ 1.0)  & 3.1 ($\div$ 1.0) &225 ($\div$ 1.0)  & 991 ($\div$ 1.0)  \\
4 cores  & 0.57 ($\div$ 1.2)  & 1.4 ($\div$ 2.2) & 58 ($\div$ 3.9)  & 251 ($\div$ 3.9)  \\
16 cores & 1.26 ($\div$ 0.5)  & 1.6 ($\div$ 1.9) & 20 ($\div$ 11.5) & 66 ($\div$ 14.9) \\
64 cores & --           & --         & 11  ($\div$ 20.5)  & 23 ($\div$ 42.4) \\
\hline
\end{tabular*}
\end{table}

We have implemented the algorithm from the previous sections for allowing its
application to actual biological sequences. The implementation uses C++ with
and Intel MMX assembly code. 
As input data for our tests, we used different biological sequence data sets 
and a fixed window size of~100 (the nature of the sequences does not affect the
running time of our algorithm, but may affect the impact of the heuristic
speedup employed by the heuristic method we compare the results to). In all
experiments, we used a vertical step size of~5, i.e.\ we only compare every
fifth window in the first input to all windows in the other string. Using the
scoring scheme as described in Section~\ref{sec:lcs} induces a window size
of~200 due to the constant-size blowup of the alignment dag.
For comparing the results to existing methods, we implemented an alternative
fast method for bit-parallel LCS computation~\cite{Crochemore+:01} 
to compute the pairwise alignment scores (``BLCS'') -- our 64-bit
implementation of this algorithm achieves a speedup of 32 over the
standard dynamic programming algorithm for inputs of length~200. 
Furthermore, we compared our results to the code used
in~\cite{Ott:09} (``Heur'') which uses the standard dynamic programming 
algorithm~\cite{Wagner+:74,Needleman1} and a heuristic to speed up computation 
when a minimum alignment score for a window pair is specified. In
the Sea-16, vectors of 16-bit values were used. Using the results from
Section~\ref{sec:seawinwin}, we can improve this to use 8-bit values, by
computing $(200,2)$-restricted highest-score matrices.
The results of our experiments are shown in Table~\ref{tbl:compresults}.
We see that the seaweed-based algorithm is fastest for all data sets. We also
see that the heuristic employed by Heur makes this algorithm more
effective than the BLCS method for long sequences. However, we note that
it would be possible to adapt BLCS to make use of the same heuristic
speedup. Overall, these results show that the seaweed algorithm
is highly competitive against the repeated dynamic programming approach, and
that particularly the byte vector version (Sea-8) is more than seven times
faster than the best existing method.

We further conducted experiments to study the scalability on larger numbers of
processors using MPI by distributing the computation of the strips between
multiple processors (see Table~\ref{tbl:compresults}, bottom result sets). 
We obtained good speedup especially for the large
datasets both on small and larger parallel systems. 
Note that our sample datasets are still rather small. We plan to apply the
algorithm to whole-genome comparison, which involves much larger input
sequences, and hence better speedup on more processors.

\section{Conclusions and Outlook}\label{sec:conclusions}
In this paper, we present a practical algorithm for local string comparison by edit distance
filtered dot plots which uses vector-parallelism and recent algorithmic results to achieve both improved
asymptotic cost and performance over applying optimized standard methods. We
have further shown results from a coarse-grained parallel implementation of the
algorithm, which achieved good speedup on different parallel systems.
Further performance could be gained by using SSE~\cite{Intel:09} or newer vector
architectures like Larrabee~\cite{Abrash:09} for implementing our code.
A few algorithmic improvements are possible as well. 
In~\cite{TiskinL:07:Applications}, a tree approach is proposed to avoid
recomputing all seaweeds in each strip of height $w$, which allows to perform
the computation in time $O(mn)$. We are currently investigating a 
practical variation of this theoretical method which further reduces the
dependency on the window size, and may improve the algorithm shown here.
Moreover, we believe that it is possible to use a similar heuristic to the one
applied in~\cite{Ott:09} to further improve performance.

\bibliographystyle{plain}

\end{document}